# High quality factor metasurfaces for two-dimensional wavefront manipulation

Claudio U. Hail[1], Morgan Foley[2], Ruzan Sokhoyan[1], Lior Michaeli[1], Harry A. Atwater[1]*

[1] Thomas J. Watson Laboratory of Applied Physics, California Institute of Technology, Pasadena, California 91125

[2] Department of Physics, California Institute of Technology, Pasadena, California 91125

* Correspondence and requests for materials should be addressed to H.A.A (email: haa@caltech.edu).

## Abstract

**The strong interaction of light with micro- and nanostructures plays a critical role in optical sensing, nonlinear optics, active optical devices, and quantum optics. However, for wavefront shaping, the required local control over light at a subwavelength scale limits this interaction, typically leading to low-quality-factor optical devices. Here, we demonstrate an avenue towards high-quality-factor wavefront shaping in two spatial dimensions based on all-dielectric Huygens metasurfaces by leveraging higher-order Mie resonances. We design and experimentally realize transmissive band stop filters, beam deflectors and radial lenses with measured quality factors in the range of 202–1475 at near-infrared wavelengths. The excited optical mode and resulting wavefront control are both local, allowing versatile operation with finite apertures and oblique illumination. Our results represent an improvement in quality factor by nearly two orders of magnitude over previous localized mode designs, and provide a design approach for a new class of compact optical devices.**

The recirculation of light in a confined optical mode is a ubiquitous method to amplify the interaction of light and matter. In this respect, the ability to confine the light to the resonating mode is quantified by the quality factor, $Q$, as the energy stored per round-trip optical loss in the resonator. With optical micro- and nanostructures including Fabry-Pérot cavities[1,2], whispering gallery mode resonators[3–6], photonic crystals[7,8], guided mode structures[9], and bound states in the continuum (BIC)[10–12], quality factors of up to $10^8$ have been demonstrated. The high level of field enhancement and confinement attained in these structures has led to many advances in sensing[13,14] active optical devices[15,16], light sources[7,17] and amplification of photon-matter coupling[2,18]. However, in general, as the mode volume of an optical resonator decreases and the mode becomes more localized, more radiative decay channels become available and the field enhancement and quality factor diminish[19]. As a result, there is typically a tradeoff between spatial mode localization and the attainable quality factor.

In optical metasurfaces, a subwavelength-spaced array of localized resonators are used to abruptly manipulate the phase, amplitude, polarization, and spectrum of light at an interface[20]. These structures hold promise to revolutionize many areas of optical imaging, communication, sensing, and display technology. Numerous optical components and phenomena have been realized using metasurfaces such as efficient flat lenses[21–23], on-chip holography[24,25] and dynamic beam steering[26,27]. Attaining strong light matter interaction, and hence high quality factors, in metasurfaces is particularly desirable, as it enables the realization of efficient active wavefront manipulation, nonlinear parametric conversion, highly responsive optical sensing and tailored light emission. However, the required subwavelength scale wavefront control, imposes a limit on the resonator size, leading to significant radiative loss. As a result, most Huygens metasurfaces have been broadband and have relied on dielectric structures with limited light confinement and hence quality factor ($Q < 15$)[22,23,28]. Only recently, advances in wavefront manipulation with increased quality factors have been made with structures relying on extended guided mode resonance[29–31] and nonlocal modes based on bound states in the continuum[32,33]. Notably, achieving simultaneous local control over a wavefront with resonance phase and high quality factor remains an outstanding challenge[34,35]. Here, we demonstrate an avenue towards high quality factor metasurfaces based on Huygens metasurfaces that leverage higher-order Mie resonances to locally manipulate the wavefront of light in two dimensions based on resonance phase.

**Results**

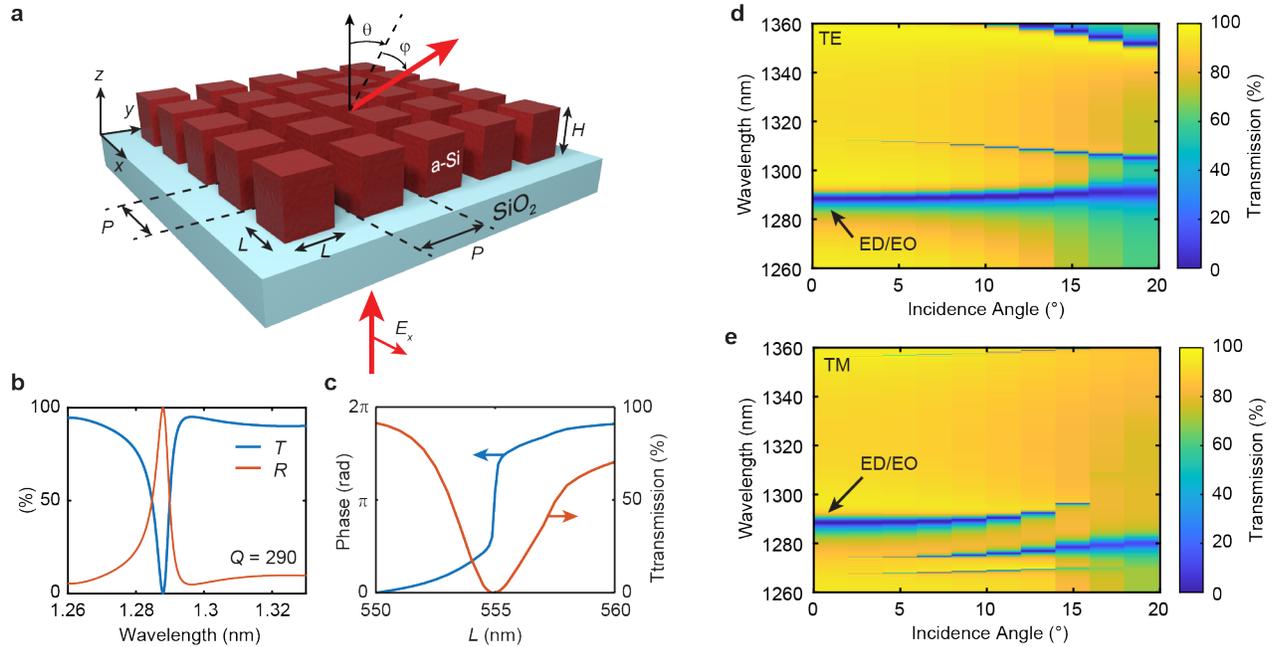

**Figure 1. | High quality factor metasurfaces for two-dimensional wavefront manipulation. a,** Schematic of the metasurface consisting of amorphous silicon nanoblocks on a glass substrate. The surface is illuminated at normal incidence and can deflect light along the angles θ and φ in the full upper hemisphere. **b,** Calculated transmission (T) and reflection (R) spectra of the metasurface with $P$ = 736 nm, $H$ = 695 nm, $L$ = 555 nm. **c,** Calculated transmission intensity and phase with varying nanoblock side length $L$, $P$ = 736 nm, and $H$ = 695 nm at a wavelength of λ = 1288 nm. **d, e,** Simulated transmission of the metasurface in (**a**) with varying angle of incidence for TE (**d**) and TM (**e**) polarization.

Figure 1a illustrates our high quality factor optical metasurface for wavefront manipulation in two dimensions. The surface consists of subwavelength-spaced amorphous silicon nanoblocks of length $L$ and height $H$ arranged in a square array with periodicity $P$ on a transparent glass substrate. The structure dimensions are in the sub-diffractive regime, both in air and in the substrate, to avoid exciting any lattice modes. The geometric parameters are chosen to induce and spectrally overlap an electric dipole (ED) and electric octupole (EO) mode in the nanoblocks at near-infrared wavelengths. With the ED and EO mode, the surface operates as a higher-order Huygens metasurface enabling local control over the transmitted wavefront. Figure 1b shows the transmission and reflection spectrum of the metasurface with a uniform nanoblock side length $L$ = 555 nm, as calculated with finite difference time domain (FDTD) simulations (see Methods for details). A multipole expansion of the modes inside a single nanoblock embedded in the array shows that the ED and EO modes are in phase and of similar strength on resonance (see Supplementary Note 1)[36,37]. The destructive interference of the ED/EO mode and the illumination induces a sharp dip in transmission and near-unity reflection on resonance. A quality factor of $Q$ = 290 is obtained from a Fano resonance fit[38]. On resonance, the electric field in the resonator shows a strong field enhancement of more than 28 times the incident field amplitude (see electric and magnetic field profiles Supplementary Note 1). Furthermore, the EO mode is clearly visible in the electric and magnetic field profiles.

Due to the near-field coupling of the ED and EO among neighboring elements, the resonance is dependent on the array size, i.e., number of repetitions of the unit cell, which is the signature of a partially delocalized mode. This array size dependence is similar to that seen in lower-order Huygens metasurfaces[28] or asymmetry-induced quasi-BIC structures[39]. Our calculations suggest that for an array size beyond 10 × 10 unit cells, there is no further significant change in the modal properties (see Supplementary Fig. 1). To further provide insight into the ED/EO mode we study its degree of localization. Exciting the mode in a single nanoblock of the array shows that it localized within the nanoblock but extends primarily to its nearest neighbors along the direction of polarization (see Supplementary Note 2). From this calculation, we determine a mode volume of 3.6 times the volume of a unit cell, or 0.86 $\lambda^3$, where $\lambda$ is the wavelength in free space. The mode localization can also be inferred from its dependence on the incidence-angle of the illumination (see Fig. 1d and e). The ED/EO modes can be excited at oblique incidence for both TE and TM polarization. We observe a spectral shift of the resonance of less than 2 nm for a 10° change in the incident angle. For comparison, a fully delocalized lattice mode or guided mode resonance at the same resonance wavelength would result in a spectral shift on the order of 130 nm for the same 10° change[9]. The observed flat angular dispersion of our surface confirms that the studied high-Q mode is localized. Furthermore, the dispersion is of similar order as reported in lower-order Huygens metasurfaces based on the spectral overlap of an electric and magnetic dipole[40,41].

The local response of our higher-order Huygens metasurface enables local wavefront manipulation by modulating the phase of the transmitted light. Varying the length *L* of the nanoblocks, spectrally shifts the ED and EO modes, and hence the resonance wavelength. The spectral shift accompanying the change in length *L* allows employing the resonance phase to impose a phase shift on the transmitted light. Figure 1c shows that varying the nanoblock side length by only ± 1%, allows the phase of the transmitted light to be controlled over almost the entire 0–2π range at a fixed wavelength of λ = 1288 nm. The inherent symmetry of the unit cell and the resonant mode also result in a polarization-independent response. Most remarkably, a small variation (≤ 1%) of the length *L* by *dL* of one block in an array of nanoblocks with uniform length *L*, manifests itself in a local resonance shift of the single nanoblock acting as an effective point source scatterer (see Supplementary Note 2). Furthermore, the resonant mode remains intact, and the high quality factor is preserved. These characteristics render our metasurface, which exhibits four-fold rotational symmetry and supports partially delocalized ED/EO modes, uniquely suitable for high quality factor wavefront manipulation in two dimensions.

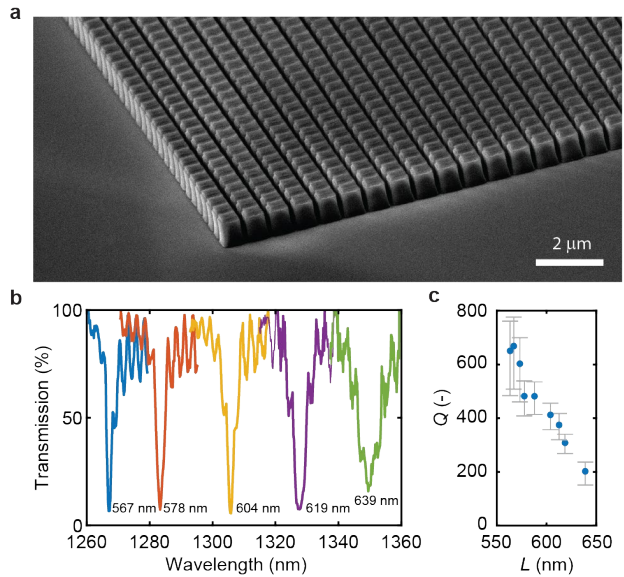

**Figure 2. | High quality factor resonances in silicon nanoblock arrays. a,** Scanning electron micrograph of a metasurface with uniform nanoblock side lengths. **b,** Experimentally measured transmission spectra of high quality factor metasurfaces with varying nanoblock side length. The measured nanoblock side length in nm is indicated next to each curve. Here, $P$ = 736 nm and H = 695 nm. **c,** Experimentally measured quality factors with varying nanoblock side lengths as determined by a Fano lineshape fit. The error bars illustrate the 95% confidence interval of the fit.

We fabricated our high-Q metasurfaces by single-step electron beam lithography and dry etching of plasma-deposited amorphous silicon on a glass substrate (see Methods for details). Figure 2a illustrates a scanning electron micrograph of a metasurface with uniform nanoblock side lengths. To experimentally characterize the surfaces, we employed linearly polarized, normally incident light through the substrate from a wavelength-tunable diode laser, collected the transmitted light with an objective lens and imaged it onto an InGaAs IR-camera, or focused it onto a power meter (see Methods for details). Figure 2b shows the measured transmission spectra of metasurfaces with uniform nanoblock side lengths. A strong dip in transmission is observed on resonance with a narrow linewidth ranging between 2–8 nm. On resonance the measured minimum transmission ranges between 6–16%. The corresponding measured quality factors range between 202–668 as determined by fitting a Fano lineshape function. With uniform nanoblock sizes, this metasurface acts as an ultra-thin narrowband bandstop filter. An increase in nanoblock length of only a few nanometers significantly red shifts the resonance by more than the linewidth. The high-frequency oscillation in the transmission spectrum is due to the interference of light reflecting at the top and bottom surface of the substrate. The measurements here are taken with an illumination spot diameter of 30 μm. This confirms that both the finite array size, and varying incident angles have a negligible effect on the optical response of the surface. Notably, the measured quality factors are higher than expected from simulations (see Fig.1). However, when fabrication imperfections, such as non-vertical sidewalls and a structure undercut, are accounted for in the simulation, the calculated quality factors increase and good agreement with the measured transmission spectrum is obtained (see Supplementary Fig. 2). The measured quality factors of our uniformly-sized metasurfaces are higher or on par with measured quality factors of metasurfaces based on asymmetry-induced quasi-BIC[11,14,39], toroidal modes[42] or electromagnetically induced transparency[43]. However, unlike these other designs, our metasurfaces allow for local manipulation of the resonance phase. The measured near-complete extinction of the transmission on resonance is evidence that the nanoblocks are fabricated with a high degree of size uniformity. A comparison with electromagnetic simulations suggests the nanoblocks are fabricated with a standard deviation in the length $L$ of less than 6 Å (see Supplementary Fig. 3). For silicon, this represents a near single atomic layer precision in nanostructuring of the metasurface.

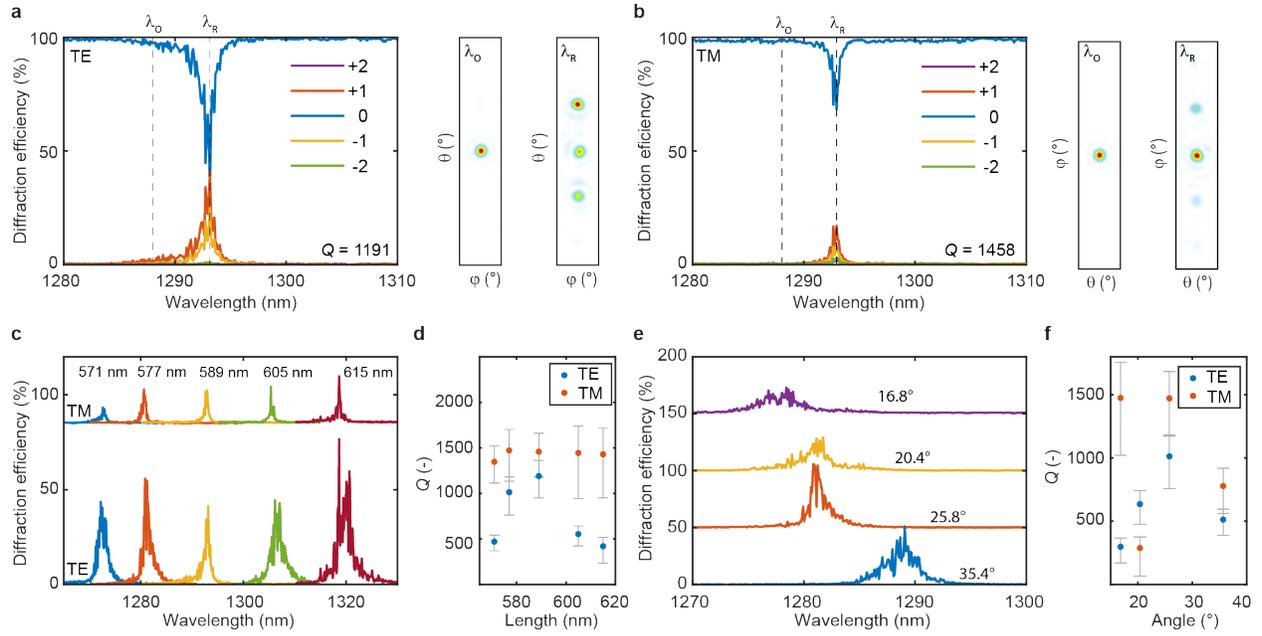

**Figure 3. | High quality factor beam deflection along two dimensions. a, b,** Experimentally measured diffraction efficiencies of the -2, -1, 0, +1 and +2 diffraction orders and Fourier plane images of a metasurface showing (**a**) TE deflection of *x*-polarized light along the *y* direction and (**b**) TM deflection of *x*-polarized light along the *x* direction. The desired diffraction order is +1, with $\theta = 26°$ and $\varphi = 26°$ respectively. **c,** Measured diffraction efficiency of TE and TM light deflection in the +1 diffraction order with varying average nanoblock side lengths. The values for TM deflection are shifted vertically by 85% for better illustration. **d,** Measured quality factors of light deflection with varying average nanoblock side length. **e,** Measured diffraction efficiency of TE light deflection with varying deflection angle. The curves are shifted vertically by 50% from each other for better visibility **f,** Measured quality factors of light deflection with varying deflection angles. The error bars in (**d**) and (**f**) illustrate the 95% confidence interval of the fit.

To selectively deflect light over a narrow wavelength range, we imprinted a linear phase gradient on the transmitted light by varying the nanoblock side lengths along one of the metasurface in-plane directions. We performed Fourier plane imaging of the transmitted light to characterize the light deflection of the metasurface. Figure 3a shows the measured spectral diffraction efficiency and images of the Fourier plane of the light transmitted through a high-Q beam deflector metasurface in the on- and off-resonance case for deflecting *x*-polarized light along the *y* direction. This results in a transverse electric (TE) polarization of the deflected light, referred to as TE deflection. On resonance, light is preferentially deflected to an angle of $\theta = 26°$ from the surface normal as determined by the linear phase gradient of the metasurface. At the design wavelength $\lambda_R$ = 1293 nm, a diffraction efficiency of 41.2% is attained. The remaining power is coupled to the normal and opposite direction at $\theta = -26°$. Notably, in a representative off-resonant case, at $\lambda_O$ = 1288 nm, 97.4% of the transmitted light remains in the surface normal direction. The measured quality factor of the TE light deflection is $Q$ = 1191. By imprinting a phase gradient along the orthogonal direction of the metasurface, light of the same polarization is deflected along the *x*-direction to an angle $\varphi = 26°$ with a maximum diffraction efficiency of 17.1% at $\lambda_R$ = 1293 nm (see Fig. 3b), due to the polarization-independent response. The result is a transverse magnetic (TM) polarization of the deflected light, referred to as TM deflection. Here, a larger quality factor of $Q$ = 1458 is measured. Off-resonance, at $\lambda_O$ = 1289 nm, 98.1% of the transmitted light remains in the normal direction. Two additional representative measurements of TE and TM deflection are illustrated in Supplementary Fig. 4 and 5.

The operating wavelength for beam deflection can be adjusted by shifting the average length of nanoblocks. Figure 3c and d illustrate the measured spectral diffraction efficiency and quality factors for TE and TM deflection for surfaces with varying operating wavelength and fixed phase gradient. By modifying the phase gradient imprinted on the surface, light can be deflected to different angles. Figure 3e and f illustrate the measured spectral diffraction efficiency for TE light deflection and quality factors for TE and TM deflection for surfaces deflecting light to different angles (the corresponding TM diffraction efficiency is shown in Supplementary Fig. 6). Overall, the obtained diffraction efficiencies on resonance for the desired deflection angle range between 22–76.6% and 4–19.1% for the TE and TM deflection, respectively. The lower diffraction efficiency obtained for TM polarized deflection compared to TE is likely due to nearest-neighbor coupling between the nanoblocks. For TM deflection, the nearest-neighbor coupling occurs along the direction of the phase gradient, therefore making this case more prone to local phase errors that arise from coupling between nanoblocks and fabrication imperfections. The attained $Q$ factors for TE and TM deflection range between $Q_{TE}$ = 298–1191 and $Q_{TM}$ = 288–1475. These results illustrate that light can be deflected along two dimensions to arbitrary angles θ and φ with high quality factor. This contrasts with previous demonstrations of high quality factor metasurfaces based on guided mode resonance structures, which are inherently one-dimensional in their light deflection capability, and require large illumination apertures and precise configuration of the incidence angle[29,30].

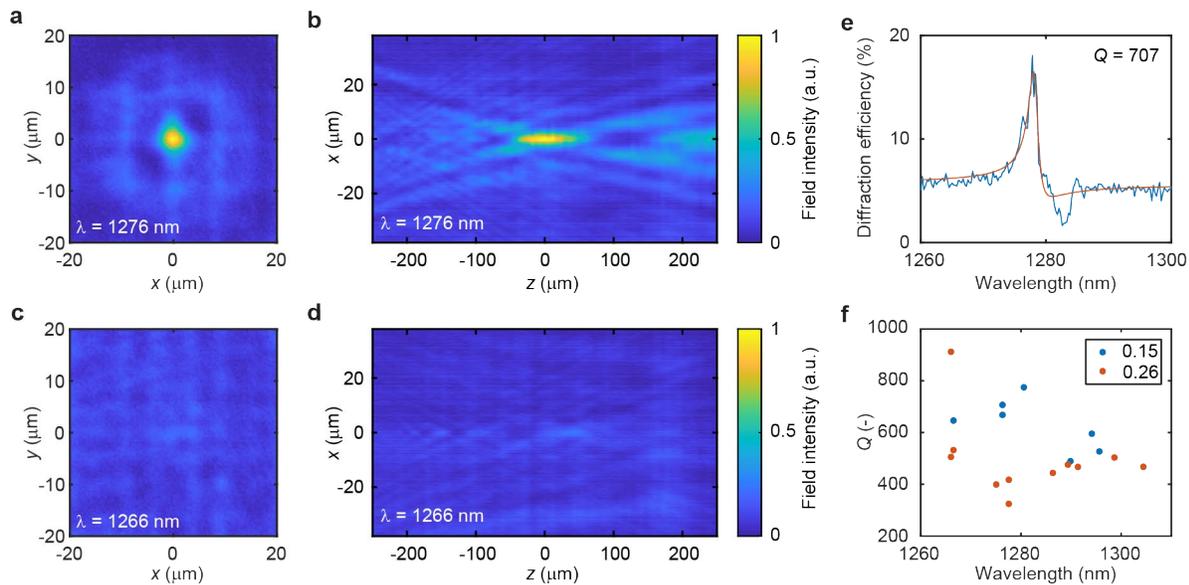

**Figure 4. | High quality factor radial metalenses for focusing along two dimensions. a,** Measured field intensity at the focal plane (*x-y* plane) on resonance at λ = 1276 nm of a high-Q metalens with 0.15 NA. **b,** Measured field intensity along the optical axis in the *x-z* plane on resonance. **c,** Measured field intensity at the focal plane off resonance at a representative wavelength λ = 1266 nm. **d,** Measured field intensity along the optical axis in the *x-z* plane off resonance. The scaling of the color maps in (**a**), (**b**), (**c**) and (**d**) are identical. **e,** Measured spectral diffraction efficiency of the metalens with 0.15 NA and $Q$ = 707. **f,** Measured quality factors of lenses with different resonance wavelengths and numerical apertures of 0.15 and 0.26.

To demonstrate the wavefront shaping capabilities of our metasurfaces for, we realize high quality factor radial metalenses that focus light along two dimensions over a narrow wavelength range. Figure 4a and b illustrate the measured electric field intensity in the focal plane and in a cross section along the optical axis of a metalens with a numerical aperture of 0.15 at the resonance wavelength of $\lambda$ = 1276 nm. On resonance, light is symmetrically focused to a near-diffraction-limited focal spot. However, at $\lambda$ = 1266 nm, at a wavelength only 10 nm away from the resonance wavelength, light propagates through the metalens without focusing (see Fig. 4c and d). Figure 4e, shows the measured spectral diffraction efficiency of the metalens, highlighting its wavelength-selective operation. The maximum diffraction efficiency is 18% on resonance. From the spectral diffraction efficiency, a quality factor of $Q$ = 707 is determined for the metalens. Figure 4f illustrates measured quality factors from different lenses with varying operating wavelength and numerical aperture. A similar trend in increased quality factors towards shorter wavelength, as in Fig. 2b, is observed. The highest quality factor $Q$ = 912 is obtained for a lens with 0.26 NA (see Supplementary Fig. 7). To our knowledge, this represents the highest quality factor radial lens demonstrated to date, and is one order of magnitude higher than previous demonstrations with nonlocal metasurfaces[33]. The Strehl ratios of the characterized lenses range within 0.25–0.53 (see Supplementary Fig. 8). We note that the performance of the metalenses demonstrated here is currently limited by fabrication accuracy. Furthermore, the coupling between neighboring elements is not optimized in the design, but is known to have a significant effect[40].

In summary, we have demonstrated the concept of a higher-order Huygens metasurface as a pathway towards high quality factor two-dimensional wavefront manipulation. By spectrally overlapping an electric dipole and electric octupole mode, a sharp resonance in transmission is obtained that enables local control of the wavefront of light. Using local phase control, we realize beam deflectors with high directivity and quality factors of $Q$ = 288–1475. We further demonstrate radial lensing with near-diffraction-limited focusing and quality factors of $Q$ = 350–912. Notably, due to the local nature of these modes, high quality factors are observed even with small illumination areas, here 30 μm in diameter. This is in contrast to recent work with guided mode resonance structures, where illumination spots of up to 500 μm are required to attain considerable quality factors[29]. The demonstrated quality factors are currently limited by fabrication imperfections and coupling between neighboring structures. With higher fabrication uniformity and lower surface roughness, we expect to attain even higher quality factors. By accounting for nanoblock coupling in the design[44] or employing inverse design concepts[45], higher diffraction efficiencies could be attained.

The high level of wavefront control and strong light interactions with our structure, as evidenced by the high quality factor, make our higher-order Huygens metasurfaces highly suitable for optical sensing[14], nonlinear optics[46], directional lasing[47] and active wavefront manipulation[48]. Our design uses resonant high-Q phase control for wavefront shaping, as opposed to previous work that has used geometric phase[32,33], and is also advantageous for implementation of active optical devices that dynamically and modulate the dielectric environment of a metasurface unit cell. For example, by introducing a refractive index change in the nanoblocks reported here using either the thermo-optic or electro-optic effect, our metasurface may be used to dynamically steer light[49]. The high-Q metasurfaces demonstrated here may also be realized at visible wavelengths, where the narrow spectral response is advantageous for display and coloring applications[50,51]. We envision that the extraordinary characteristics of higher-order Mie resonances in high-index nanoblock arrays demonstrated here will lead to numerous applications, as well as new physics[52].

**Methods**

**Experiment** The fabricated metasurfaces were characterized on a home-built optical transmission microscope (schematically illustrated in Supplementary Fig. 9). Coherent light from a wavelength-tunable diode laser (Santec TSL-510) was loosely focused on the metasurface. The transmitted light was collected with an objective lens (20x, 0.4 NA, Mitutoyo) and projected either onto a InGaAs IR camera (Xenics Bobcat 320) or a power meter (Thorlabs S122C). For measuring the transmission spectrum, the laser wavelength was scanned, and the transmitted power was recorded with the power meter. For the power normalization the sample was removed, and the illuminated power through the same area was recorded. For the beam deflection measurements, a Fourier plane was imaged onto the camera and a 0.9 NA objective lens was used to capture all diffraction orders. For characterizing the lenses, the focal plane was imaged on to the camera and a scan along the optical axis was obtained by moving the surface along the $z$ direction.

Scanning electron micrographs were acquired on an FEI Nova 200 NanoLab system to measure the sizes of the fabricated structures. For imaging, the surfaces were covered with a 2 nm thick gold layer by sputter deposition.

**Fabrication** The metasurfaces were fabricated on borosilicate glass substrates ($n$ = 1.503) with a thickness of 220 μm. To remove organic residues from the surface, the substrates were cleaned in an ultrasonic bath in acetone, isopropyl alcohol, and deionized water each for 15 min, dried using a $N_2$ gun and subsequently cleaned using oxygen plasma. Amorphous silicon was deposited onto the glass using plasma-enhanced chemical vapor deposition. In a subsequent step, the nanoblocks were written in a spin coated MaN-2403 resist layer by standard electron beam lithography. The nanoblocks were then transferred to the amorphous silicon using an $SiO_2$ hard mask with chlorine-based inductively coupled reactive ion etching. As a last step, the residual mask was removed by immersing the samples in buffered hydrofluoric acid (1:7) for 5 s, and subsequent rinsing in deionized water.

**Simulation** The numerical modelling of the nanostructures was carried out using an FDTD method. Simulations were performed with a commercially available FDTD software (Lumerical FDTD Solutions). A constant refractive index of $n$ = 1.503 was used for the borosilicate glass and a constant value of $n$ = 3.45 for amorphous silicon, as experimentally determined by ellipsometry. The simulations were carried out with a spatially coherent plane wave illumination and periodic boundary conditions were applied on all sides of the computational domain unless otherwise noted. A smallest mesh-refinement of 5 nm was used. For the oblique illumination simulations, the broadband fixed angle technique (BFAST) is used.

**Acknowledgements**

This work was supported by the Air Force Office of Scientific Research under the Meta-Imaging MURI grant #FA9550-21-1-0312 and under grant FA9550-18-1-0354. C.U.H. acknowledges support from the Swiss National Science Foundation through the Early Postdoc Mobility Fellowship grant #P2EZP2_191880. L.M. acknowledges support from the Fulbright Fellowship program and the Breakthrough Foundation. We gratefully acknowledge the critical support and infrastructure provided for this work by The Kavli Nanoscience Institute at Caltech.

**Contributions**

C.U.H, H.A.A. and R.S. conceived the project. C.U.H preformed the simulations, fabricated the devices, built the experiment, performed the measurements, and analyzed the results. R.S. and C.U.H conceived the metasurface design. M.F. assisted with simulations and fabrication. L. M. assisted with analyzing the results. C.U.H. wrote the manuscript with input from all other authors. H.A.A supervised all aspects of the project.

**Competing interests**

The authors declare no competing interests.

**Correspondence and requests for materials** should be addressed to H.A.A.

**Data availability**

The data that support the plots within this paper and other findings of this study are available from the corresponding authors upon reasonable request.